\documentclass[%
 reprint,
 amsmath,amssymb,
 aps,
]{revtex4-2}

\usepackage{graphicx}
\usepackage{dcolumn}
\usepackage{bm}

\begin{document}

\preprint{APS/123-QED}

\title{Ultrasensitive optomechanical detection of an axion-mediated force based on a sharp peak emerging in probe absorption spectrum
}%

\author{Lei Chen}
 \altaffiliation[Also at ]{Key Laboratory of Artificial Structures and Quantum Control (Ministry of Education), School of Physics and Astronomy, Shanghai Jiao Tong University, 800 Dong Chuan Road, Shanghai 200240, China}
 \author{Jian Liu}
 \altaffiliation[Also at ]{Key Laboratory of Artificial Structures and Quantum Control (Ministry of Education), School of Physics and Astronomy, Shanghai Jiao Tong University, 800 Dong Chuan Road, Shanghai 200240, China}
\author{Kadi Zhu}%
 \email{zhukadi@sjtu.edu.cn}
\affiliation{%
 Key Laboratory of Artificial Structures and Quantum Control (Ministry of Education), School of Physics and Astronomy, Shanghai Jiao Tong University, 800 Dong Chuan Road, Shanghai 200240, China 
}%




\date{\today}

\begin{abstract}
Axion remains the most convincing solution to the strong-CP problem and a well-motivated dark matter candidate, causing the search for axions and axion-like particles(ALPs) to attract attention continually. The exchange of such particles may cause anomalous spin-dependent forces, inspiring many laboratory ALP searching experiments based on the detection of macroscopic monopole-dipole interactions between polarized electrons/nucleons and unpolarized nucleons. Since there is no exact proof of the existence of these interactions, to detect them is still of great significance. In the present paper, we study the electron-neucleon monopole-dipole interaction with a new method, in which a hybrid spin-nanocantilever optomechanical system consisting of a nitrogen-vacancy(NV) center and a nanocantilever resonator is used.  With a static magnetic field and a pump microwave beam and a probe microwave beam applied, a probe absorption spectrum could be obtained. Through specific peaks appearing in the spectrum, we can identify this monopole-dipole interaction. And we also provide a prospective constraint to  constrain the interaction. Furthermore, because our method can also be applied to the detection of some other spin-dependent interactions, this work provides new ideas for the experimental searches of the anomalous spin-dependent interactions.
\end{abstract}

\maketitle


\section{\label{sec:level1} Introduction}
Axion is a new light pseudoscalar particle predicted in 1978 \cite{PhysRevLett.40.223, PhysRevLett.40.279}. Since then, it remains the most compelling solution to the strong-CP problem in QCD and a well-motivated dark matter candidate \cite{doi:10.1146/annurev-nucl-102014-022120,beringer2012review, tanabashi2018review}. Due to this, a host of ultrasensitive experiments have been conducted to search for axions and axion-like particles (ALPs) \cite{doi:10.1146/annurev-nucl-102014-022120,beringer2012review,ficek2019constraining,safronova2018search}. The exchange of such particles may cause spin-dependent forces, the framework of which was introduced by Moody and Wilczek \cite{PhysRevD.30.130} and extended by Dobrescu and Mocioiu \cite{Dobrescu_2006}. Also, some errors and omissions in \cite{PhysRevD.30.130}  and \cite{Dobrescu_2006}  are corrected in the recent papers \cite{PhysRevA.99.022113} and \cite{daido2017sign}. Here we focus on one type of spin-dependent forces: the so-called monopole-dipole interaction. Lots of laboratory ALP searching experiments \cite{wineland1991search,youdin1996limits,heckel2008preferred,terrano2015short,hoedl2011improved,petukhov2010polarized,ni1999search,jin2013direct,rong2018searching}  based on the detection of this interaction between polarized electrons/nucleons and unpolarized nucleons have been accomplished and many constraints have been established. However, the exotic monopole-dipole interction has not been observed so far. Thus it is still desirable for us to develop new methods or more advanced technologies to search this interaction.

In this paper, we propose a quantum optical method using a hybrid spin-nanocantilever quantum device to investigate this interaction between polarized electrons and unpolarized nucleons. With a pump microwave beam and a probe microwave beam applied,
we could obtain a probe absorption spectrum which contains information of the exotic interaction. We present our numerical results about the absorption spectrum. Then Based on it we demonstrate our detection method and set an estimated constraint for the coupling constants $g_s^N g_p^e$. Finally, we expect our work could enrich the methods of the experimental searches for the hypothetical interactions.
\section{\label{sec:level1} Theoretical model}
\begin{figure}
\resizebox{0.5\textwidth}{!}{%
  \includegraphics{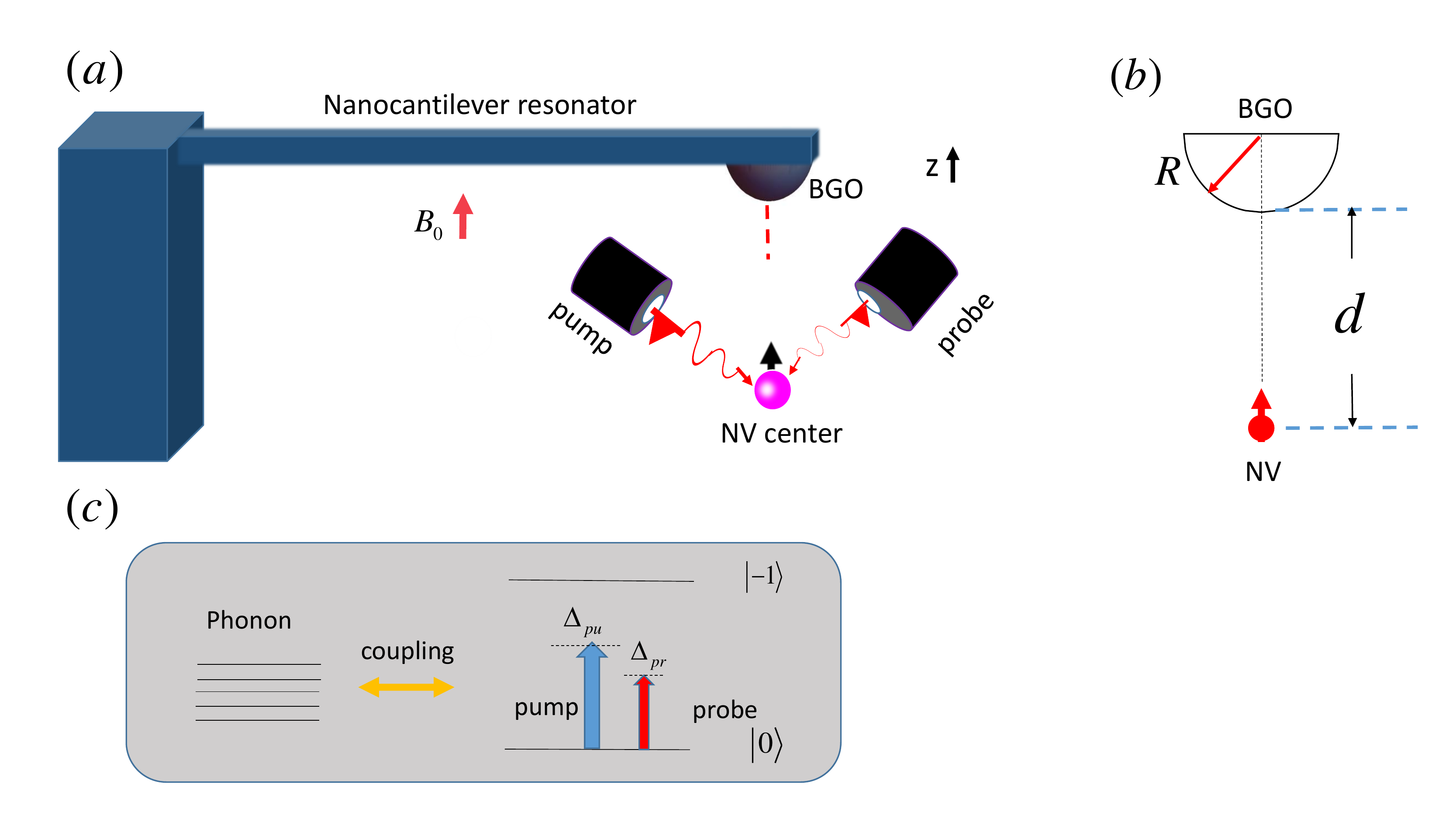}
}
\caption{Setup and the energy-level diagram. (a) Schematic setup. A $Bi_4 Ge_3 O_{12}$ half ball denoted as BGO, is placed on a nanocantilever resonator. A NV center, whose symmetry axes assumed as the z direction, is positioned under BGO. Besides, The symmetry axes of the two coincide with each other. A pump beam and a probe beam are applied to the NV center simultaneously. A static magnetic field $B_0$  is applied along the z direction. (b) The radius of BGO is R. The distance between NV and the bottom of BGO is d. (c) An energy-level diagram of the NV center spin coupled to the resonator. $ \Delta_{pu}$ and $ \Delta_{pr}$ are pump-spin detuning and probe-spin detuning respectively.}
\label{fig:1}       
\end{figure}
Here we consider a system illustrated in Fig. 1(a), which consists of a nanocantilever resonator, a $Bi_4 Ge_3 O_{12}$  half ball , and a near-surface NV center in diamond. The half ball, whose radius can be assumed as $R=25nm$ \cite{aldica2012investigations}, is denoted as BGO and placed on the resonator. The NV center, which is about 10nm close to the surface of the diamond, is positioned at a distance $d\approx25nm$ under the bottom of BGO. Furthermore, the symmetry axes of the NV center and BGO, which are both in the z direction, coincide with each other (see Fig. 1(a),(b)). Since the NV is a single electron spin and BGO is a source of unpolarized nucleons, it is assumed that there is an axion-mediated monopole-dipole interaction between this electron spin and neucleons, which can be described as \cite{Dobrescu_2006,rong2018searching} 
\begin{equation}
V_{en}(\overrightarrow r)=\frac{{\hbar^2}{g_s^N}{g_p^e}}{8 \pi m}(\frac 1{\lambda r}+\frac 1{r^2}) e^{-{\frac r\lambda}}{\overrightarrow \sigma}\cdot {\overrightarrow e},
\end{equation}
where $\overrightarrow r$  is the displacement vector pointing from the nucleon to the electron, $r=|\overrightarrow r|$, $\overrightarrow e={\overrightarrow r}/r$,  $g_s^N$ and $g_p^e$ are the scalar and pseudoscalar coupling constants of the ALP to the nucleon and to the electron, $m$ is mass of the electron, $\lambda={\hbar}/{m_a c}$ is the force range, $m_a$ is the mass of the ALP, $c$ is the speed of the light, and $\overrightarrow \sigma$ is the Pauli vector of the electron spin. The monopole-dipole interaction between all the nucleons in BGO and the electron spin is equivalent to the Hamiltonian of the electron spin in an effective magnetic field $\overrightarrow B=B {\overrightarrow {e_z}}$, where   ${\overrightarrow {e_z}}$ is the unit displacement vector along the symmetry axis of BGO (the inverse of z direction) and $B$ satisfies \cite{rong2018searching} 
\begin{equation}
 B=\frac{\hbar g_s^Ng_p^e\rho}{2m\gamma}  f(\lambda, R, d),
\end{equation}
 where $\rho=4.29 \times {10^{30}} {m^{-3}}$ is the number density of nucleons in BGO\cite{rong2018searching} , $\gamma= e/m $ is the gyromagnetic ratio of the electron spin of the NV center, $e\approx 1.6\times 10^{-19} C$ is an electron charge, and
\begin{align*}
 f(\lambda,R,d)=&\lambda [{\frac R{d+R}} e^{-\frac d \lambda}-e^{-\frac {d+R} \lambda}+e^{-\frac{\sqrt {R^2+(d+R)^2}} {\lambda}} \\ &+\frac{\lambda \sqrt{R^2+(d+R)^2}}{(d+R)^2} e^{-\frac{\sqrt{R^2+(d+R)^2}}\lambda}\\&-{\frac{\lambda d}{(d+R)^2}}e^{-\frac d \lambda} +\frac{\lambda^2}{(d+R)^2} e^{-\frac{\sqrt{R^2+(d+R)^2}}\lambda}\\ &-\frac{\lambda^2}{(d+R)^2} e^{-\frac d \lambda}].
\end{align*}

Now we demonstrate how the system works. The ground state of the NV center is an $S=1$ spin triplet with three substates $|m_s=0\rangle$  and $|m_s=\pm\rangle$ . This three substates are separated by a zero-field splitting of $\omega_s/2\pi \simeq 2.88GHz$. Applying a moderate static magnetic field $B_0 =300G$ \cite{rong2018searching} along the z direction (see Fig. 1(a)), we remove the degeneracy of the $|m_s=\pm1\rangle$  spin states.  Then the NV spin can be restricted to a two-level subspace spanned by $|0\rangle$   and $|-1\rangle$ \cite{teissier2014strain}. As a result, the Hamiltonian of the NV center in the magnetic field $B_0$ can be written as $H_{NV}=\hbar \omega_s S_z$ . $S_z$ together with $S^{\pm}$  characterize the spin operator. Next, we demonstrate how the NV center is coupled to the resonator.

The resonator is described by the Hamiltonian $H_r=\hbar\omega_r c^+ c$ \cite{rabl2009strong}, where $\omega_r$ is the frequency of the fundamental bending mode and $c$  and $c^+$ are the corresponding annihilation and creation operators. As mentioned above, the NV electron is in an effective magnetic field $\overrightarrow B$ . Because the cantilever resonator drives BGO to vibrate, the NV electron feels an effective time-varying magnetic field ${\overrightarrow B}_{eff}(t)$ . Then when the magnetic field $B_0$ is applied, the Hamiltonian of the system could be written as \cite{rabl2009strong}:                                                                                                              \begin{align}
 H_s&=H_{NV}+H_r+\hbar g(c+c^+)S_z \notag \\
    &=\hbar\omega_s S_z+\hbar\omega_r c^+ c+\hbar g(c+c^+)S_z,
\end{align}
where the coupling coefficient $g$ satisfies:
\begin{equation}
g=\frac{g_s \mu_B G_m a_0}\hbar.
\end{equation}
Here $g_s\approx2$ is g-factor of the electron, $\mu_B=9.27\times 10^{-24} A\cdot m^2$ is the Bohr magneton, $G_m$ is the gradient of the magnetic field ${\overrightarrow B}_{eff}(t)$ at the position of the NV center, $a_0$ is the amplitude of zero-point fluctuations for the whole of the cantilever resonator and BGO and it can be described as
\begin{equation}
a_0=\sqrt{\hbar/2 m_s \omega_r},
\end{equation}
$m_s$ is the sum of the mass of the resonator and BGO. In addition, we can derive
\begin{equation}
G_m=\frac{{\hbar^2}|{g_s^N}{g_p^e}|\rho}{2m\gamma}\lambda |u|,
\end{equation}
 with
\begin{align}
u=&(\frac{R\lambda+\lambda^2}{4{R^3}}-\frac 1{2\lambda})e^{-\frac R\lambda}+{\frac 1\lambda}e^{-2\frac R\lambda} \notag \\ &-(\frac 2{\sqrt5 \lambda}+\frac 1{2R} +\frac{{\sqrt5}\lambda}{4{R^2}}+\frac {\lambda^2}{4{R^3}})e^{-\frac {{\sqrt5} R}\lambda}.
\end{align}

According to our scheme, a strong pump microwave beam and a weak microwave beam are applied to the NV center simultaneously (see Fig. 1(a)). Then the vibration mode of the resonator can be treated as phonon mode and some nonlinear optical phenomena occur. An energy level diagram of the NV center spin coupled to the cantilever resonator is illustrated in Fig. 1(c), where $ \Delta_{pu} \equiv \omega_s - \omega_{pu}  $ and $ \Delta_{pr} \equiv \omega_s -\omega_{pr} $ are pump-spin detuning and probe-spin detuning respectively.
Next we attempt to derive the expression of the first order linear optical susceptibility. The Hamiltonian of the NV center spin in $B_0$ coupled with two microwave fields reads as follows \cite{boyd2008nonlinear}:
\begin{align}
H_{int}=&-\mu(S^+ P_{pu} e^{-i\omega_{pu} t}+S^- P_{pu}^* e^{i\omega_{pu} t})\notag \\
        &-\mu(S^+ P_{pr} e^{-i\omega_{pr} t}+S^- P_{pr}^* e^{i\omega_{pr} t}),
\end{align}
where $\omega_{pu}(\omega_{pr})$ is the frequency of the pump field (probe field), $P_{pu}(P_{pr})$ is the slowly varying envelope of the pump field (probe field), and $\mu$ is the induced electric dipole moment. Consequently, when the magnetic field and two beams are applied, the Hamiltonian of the system can be described as:
\begin{align}
H=&H_s + H_{int}\notag\\
 =&\hbar \omega_s S_z +\hbar \omega_r c^+ c+\hbar g (c+c^+)S_z\notag\\
  &-\mu(S^+ P_{pu} e^{-i\omega_{pu} t}+S^- P_{pu}^* e^{i\omega_{pu} t})\notag \\
  &-\mu(S^+ P_{pr} e^{-i\omega_{pr} t}+S^- P_{pr}^* e^{i\omega_{pr} t}).
\end{align}
Then, we transform Eq. (9) into a rotating frame at the pump field frequency $\omega_{pu}$ to simplify the following solution procedure, and obtain
\begin{align}
H^\prime =&\hbar\Delta_{pu} S_z +\hbar \omega_r c^+ c +\hbar g (c+c^+)S_z\notag\\ &-\hbar(\Omega S^+ +\Omega^* S^-)\notag\\
    &-\mu(S^+ P_{pr} e^{-i\delta t}+S^- P_{pr}^* e^{i\delta t}),
\end{align}
where $\Omega=\mu P_{pu}/\hbar$ is the Rabi frequency of the pump field, and $\delta=\omega_{pr}-\omega_{pu}$ is the pump-probe detuning.

Applying the Heisenberg equations of motion for operators $S_z , S_-$ and $\zeta=c^+ +c$, introducing the corresponding damping and noise terms, we derive the three quantum Langevin equations as follows \cite{boyd2008nonlinear,gardiner2000quantum} :
\begin{align}
\frac{d S_z}{dt}=&-\nu_1 (S_z+\frac 12 )+i\Omega(S_+ -S_-)\notag \\ &+i{\frac \mu\hbar}(S^+ P_{pr} e^{-i\delta t}-S^- P_{pr}^* e^{i\delta t}),
\end{align}
\begin{align}
\frac{d S^-}{dt}=&[-\nu_2-i(\Delta_{pu} +g\zeta)]S^- -2i\Omega S_z \notag\\
&-2i{\frac \mu\hbar}S_z P_{pr} e^{-i\delta t}+\hat \Theta ,
\end{align}
\begin{equation}
\frac{d^2 \zeta}{d t^2}+\gamma_n \frac{d\zeta}{dt}+\omega^2_r \zeta=-2g\omega_r S_z+\hat\Upsilon .
\end{equation}
In Eqs. (11)-(13),  $\nu_1$ and $\nu_2$ are the electron spin relaxation rate and dephasing rate respectively. $\gamma_n$  is the decay rate of the high-Q cantilever resonator.  $\hat \Theta $ is the $\delta$-correlated  Langevin noise operator, which has zero mean $\langle \hat \Theta  \rangle=0$  and obeys the correlation function $\langle {\hat \Theta }(t){\hat \Theta }^+ (t^\prime)\rangle\sim\delta(t-t^\prime)$   . Thermal bath of Brownian and non-Morkovian process affects the motion of the resonator \cite{gardiner2000quantum, giovannetti2001phase}, the quantum effect of which will be only observed when the high quality factor $Q\gg1$. Thus, the Brownian noise operator could be modeled as Markovian with $\gamma_n$. The Brownian stochastic force satisfies $\langle \hat\Upsilon \rangle=0 $ and \cite{giovannetti2001phase}
\begin{equation}
\langle {\hat\Upsilon}^+ (t){\hat\Upsilon}(t^\prime)\rangle=\frac{\gamma_n}{\omega_{nt}}\int \frac{d\omega}{2\pi} \omega e^{-i\omega(t-t^\prime)}[1+coth(\frac{\hbar \omega}{2 K_B T})].
\end{equation}
To go beyond weak coupling, we can always rewrite each Heisenberg operator as the sum of its steady-state mean value and a small fluctuation with zero mean value as follows:
\begin{equation}
S^- =S_0^- +\delta S^- ,S_z =S_0^z +\delta S_z ,\zeta=\zeta_0 +\delta \zeta.
\end{equation}
Then, we insert these equations into the Langevin equations (11)-(13), neglecting the nonlinear term $\delta \zeta \delta S^-$  . Since the optical drives are weak, we can identify all operators with their expectation values and drop the quantum and thermal noise terms \cite{weis2010optomechanically}. Furthermore, we make the ansatz \cite{boyd2008nonlinear,weis2010optomechanically}:
\begin{gather}
\langle \delta S_z \rangle=S_+^z e^{-i \delta t}+S_-^z e^{i \delta t},\\
\langle \delta S_- \rangle=S_+ e^{-i \delta t}+S_- e^{i \delta t},\\
\langle \delta \zeta \rangle=\zeta_+  e^{-i \delta t}+\zeta_- e^{i \delta t}.
\end{gather}
Since the first order linear optical susceptibility can be described as
\begin{equation}
\chi^{(1)} (\omega_{pr})=\frac{\mu S_+}{P_{pr}},
\end{equation}
 with the ralationship of $\nu_1=2 \nu_2$  assumed \cite{boyd2008nonlinear}, we can finally obtain:
\begin{equation}
\chi^{(1)}(\omega_{pr})=\frac{\mu^2}\hbar \frac{\omega_0 K_4+K_1^* K_5}{K_5 \Omega-K_3 K_4},
\end{equation}
where
\begin{align}
K_1&=\frac{\Omega\omega_0}{i\nu_2 -\Delta_{pu} +\frac{g^2 \omega_0}{\omega_r}},\notag\\
K_2&=\Delta_{pu} +i\nu_2 -g^2 \frac{\omega_0}{\omega_r}+\delta,\notag\\
K_3&=\Delta_{pu} -i\nu_2 -g^2 \frac{\omega_0}{\omega_r}-\delta,\notag\\
K_4&=2 g^2 \eta K_1^* \Omega-2\Omega_2 +K_2 (\delta+i\nu_1),\notag\\
K_5&=2K_2(g^2 \eta K_1 -\Omega).
\end{align}
In addition, the auxiliary function $\eta$  satisfies
\begin{equation}
\eta=\frac{\omega_r}{\omega_r^2 -\delta^2 -i\delta\gamma_n},
\end{equation}
and the population inversion $\omega_0$ is determined by
\begin{equation}
(\omega_0 +1)[\nu_2^2 +(g^2 \frac{\omega_0}{\omega_r}-\Delta_{pu})^2]=-2 \Omega^2 \omega_0 .
\end{equation}

Till now, the expression of $\chi^{(1)} (\omega_{pr})$ has been derived. Then we plot the probe absorption spectrum ( probe absorption, i.e.the imaginary part of $\chi^{(1)} (\omega_{pr})$, as a function of pump-probe detuning $\delta$ ) using appropriate parameters in the following section.

\section{Detection method}
\begin{figure}
\resizebox{0.5\textwidth}{!}{%
  \includegraphics{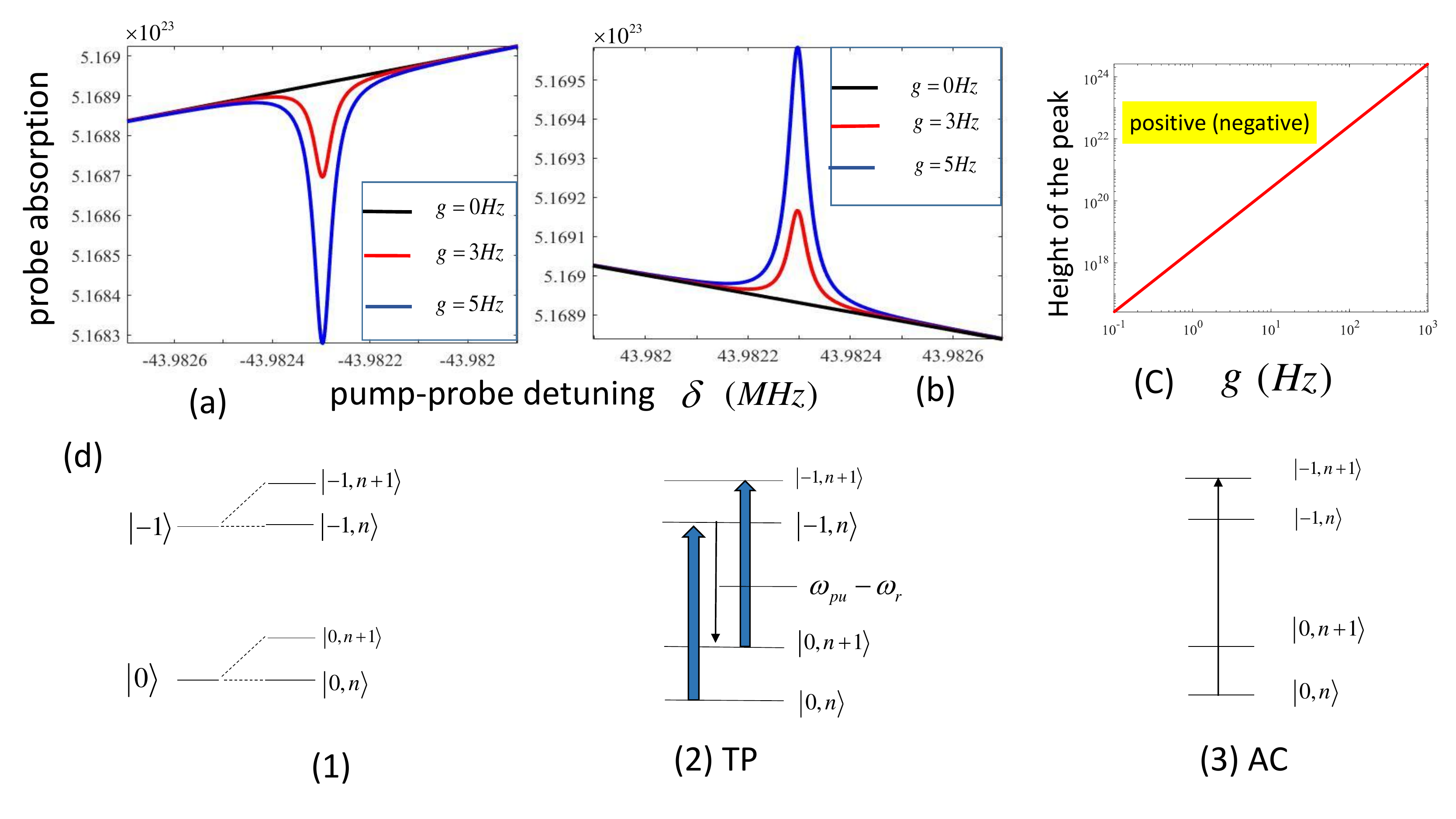}
}
\caption{Graph of  probe absorption spectrums, the relationship between height of peak and the value of $g$, and a dressed-state picture to interpret the peaks. (a)-(b) The probe absorption spectrums include one straight line and two curves, plotted using different colors around $\delta=-\omega_r$  ((a)) and $\delta=\omega_r$ ((b)), corresponding to the cases of $g=0,3,5Hz$ respectively. The parameters used are $\omega_r/2\pi\approx7MHz, {\gamma_n}/2\pi=7Hz ,\Delta_{pu}=0 ,\Omega=1KHz , \mu=10D,$ and $\nu_1 =2\nu_2 =2KHz$ . (c) The height of positive (negative) peak in the probe absorption spectrum as a function of $g$. (d) Each of the features in the above spectrums is identified by the corresponding transition between dressed states of the NV electron spin. TP denotes the three-photon resonance and AC denotes the ac-Stark-shifted resonance. }
\label{fig:2}
\end{figure}
According to Eqs. (20)-(23), using appropriate parameters, we plot the probe absorption spectrums for different values of $g$ around the points of $\delta=-\omega_r$ and $\delta=\omega_r$  in Fig. 2. These parameters are presented at first. We consider a ultraclean Si nanocantilever of dimensions (l, w, t)=(3000,50, 50) nm with a fundamental frequency of ${\omega_r}/2\pi\approx7MHz$ \cite{sidles1995rev}. Here l, w, t denote length, width, and thickness respectively. Then the amplitude of zero-point fluctuations is $a_0 \approx5\times10^{-13}m$. The quality factor Q of this Si nanocantilever resonator can reach up to ${10}^6$ in an ultralow temperature($\leq 30mk$) \cite{moser2014nanotube}. Consequently the decay rate of it is ${\gamma_n}/2\pi={\omega_r}/{2\pi Q}=7Hz$ . Furthermore, we assume $\Delta_{pu}=0$, the Rabi frequency of the pump field is $\Omega=1KHz$, and the induced electric dipole moment is $\mu=10D$. The NV electron spin dephasing time $T_2$ can be selected as $T_2 =1ms$ \cite{balasubramanian2009ultralong}. Thus the corresponding dephasing rate $\nu_2$  is ${\nu_2}=1/{T_2}=1KHz$. Since the relationship of $\nu_1 =2\nu_2$  has been assumed, the electron spin relaxation rate is $\nu_1 =2KHz$. Next, we describe the plot in Fig. 2.

In Fig. 2(a), there is a negative sharp peak centered at $\delta=-\omega_r$  in the red curve ($g=3Hz$), the rest of which coincides with the black straight line ($g=0Hz$). And it is the same for the blue curve ($g=5Hz$), except the peak of which is larger. In Fig. 2(b), a positive steep peak centered at $\delta=\omega_r$ appears for the cases of the red curve and the blue, while the rest of these two curves coincide with the black straight line. Furthermore, the blue peak is larger than the red, which is same as the situation in the left. We can also take other values of g into consideration, though not illustrated in Fig. 2. In sum, for the probe absorption spectrum of $g=0$ , there is no peak at $\delta=\pm\omega_r$ , around each of which is only a straight line. On the contrary, for a probe absorption spectrum of $g>0$  , a negative peak and a positive peak appear at $\delta=-\omega_r$  and $\delta=\omega_r$  respectively. Furthermore, when the value of g increases, both peaks become larger. In addition, we have studied the relationship between the heights of two peaks and the value of $g$.  Evidently, the heights of the positive peak and the negative one are both functions of $g$ when ${10}^{-1}Hz\le g \le {10}^{3}Hz $. And the graphs of these two functions which overlap with each other completely are plotted in Fig. 2(c).

Two peaks in a probe absorption spectrum for any positive value of $g$ can be interpreted by a dressed-state picture, in which the original energy levels of the NV electron spin $|-1\rangle$  and $|0\rangle$  have been dressed by the phonon mode of the cantilever resonator. Consequently, $|-1\rangle$  and $|0\rangle$  split into dressed states $|-1,n\rangle$  and $|0,n\rangle$  , where $|n\rangle$   denotes the number states of the phonon mode (see part (1) of Fig. 2(d)). The feature of the negative peak can be interpreted by TP, which denotes the three-photon resonance. Here the NV spin makes a transition from the lowest dressed state $|0,n\rangle$   to the highest dressed state $|-1,n+1\rangle$   by the simultaneous absorption of two pump photons and the emission of a photon at $\omega_{pu} -\omega_r$ (see part (2) of Fig. 2(d)). Meanwhile, the feature of the positive peak corresponds to the usual absorptive resonance of the NV spin as modified by the ac Stark effect, shown by part (3) of Fig. 2(d).

Based on what Fig. 2 shows, we now demonstrate our method of searching for the NV electron-nucleon monopole-dipole interaction. From Eqs. (4)-(7), it is derived that:
\begin{equation}
g=\frac{g_s \mu_B a_0 \rho}{2m\gamma}\lambda |u||g_s^N g_p^e|
\end{equation}

with $a_0=\sqrt{\hbar/2m_s \omega_r}$  , and
\begin{align*}
u=&(\frac{R\lambda+\lambda^2}{4{R^3}}-\frac 1{2\lambda})e^{-\frac R\lambda}+{\frac 1\lambda}e^{-2\frac R\lambda}  \\ &-(\frac 2{\sqrt5 \lambda}+\frac 1{2R} +\frac{{\sqrt5}\lambda}{4{R^2}}+\frac {\lambda^2}{4{R^3}})e^{-\frac {{\sqrt5} R}\lambda}.
\end{align*}
We specify the force range $\lambda$ . Consequently $u$  is determined and we assume it is not equal to zero. Then the value of $g$ is only dependent on the value of $g_s^N g_p^e$ . Thus there would be a unique probe absorption spectrum for an arbitrary value of $g_s^N g_p^e$.  When $g_s^N g_p^e =0$ , i.e., there are no monopole-dipole interaction between the NV electron spin and nucleons, $g=0$ . In this case, in the corresponding probe absorption spectrum there is no peak at $\delta=\pm\omega_r$ and only one straight line around each of two points. On the contrary, when $g_s^N g_p^e \ne0$ , $g>0$ . Consequently, a negative peak centered at $\delta=-\omega_r$  and a positive peak centered at $\delta=\omega_r$ appear in the corresponding spectrum. In addition, once the absolute value of $g_s^N g_p^e$  increases, both peaks will become larger. To sum up, if $\lambda$  is assumed and the related $u$ is not equal to zero, both the positive peak at $\delta=\omega_r$  and the negative peak at $\delta=-\omega_r$  in a probe absorption spectrum can be considered as a signature of the NV electron-nucleon monopole-dipole interaction. And a larger positive peak reflect a larger value of $|g_s^N g_p^e|$  , i.e., a stronger interaction, and so do a larger negative one. Next, we take a special case for example in which $\lambda=100nm({10}^{-7}m)$ is assumed and the corresponding $u$ is $u=-1242104.856\ne0$ .

\begin{figure}
\resizebox{0.5\textwidth}{!}{%
  \includegraphics{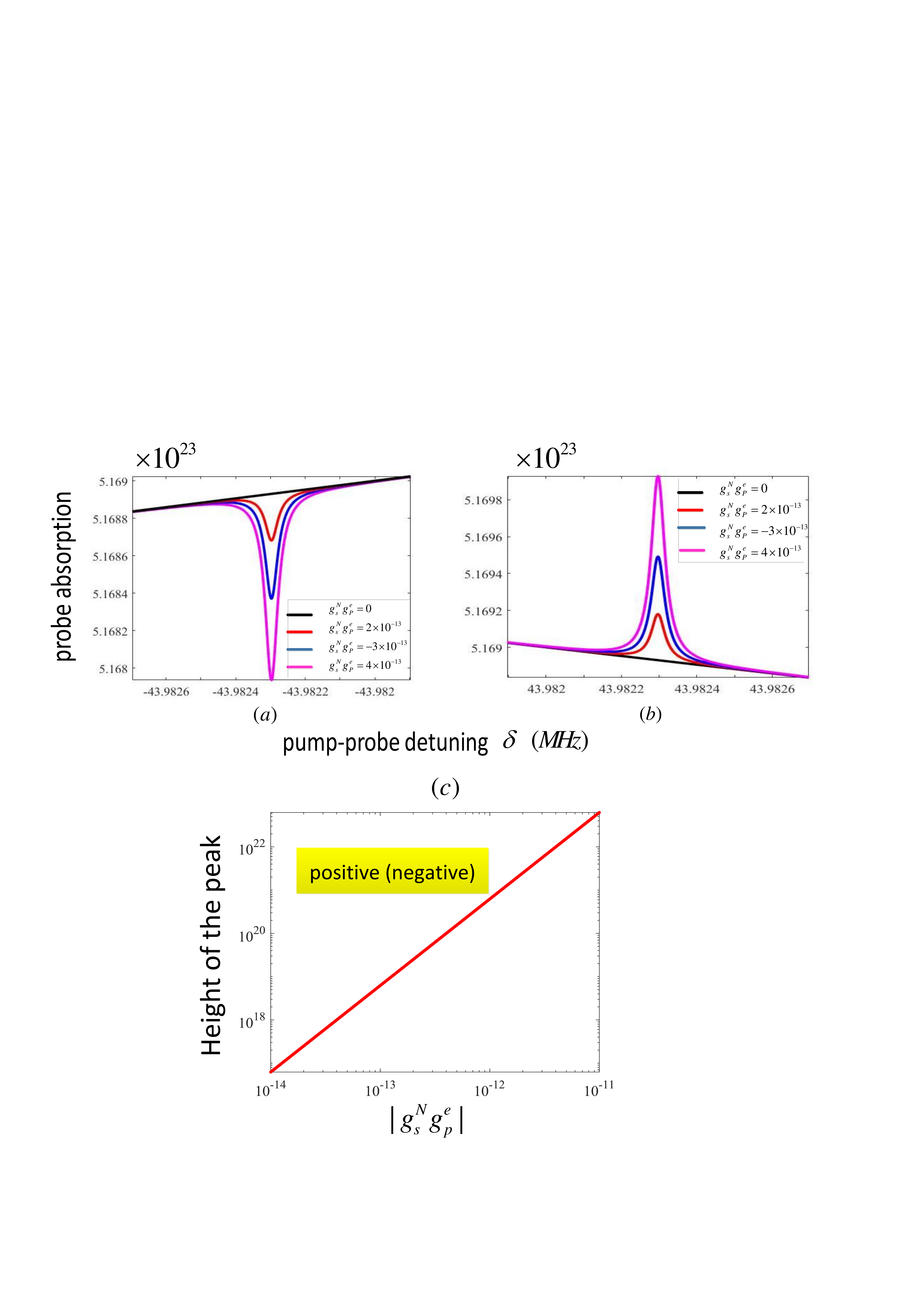}
}
\caption{Illustration of the case of $\lambda=100nm$ . (a)-(b) The probe absorption spectrums for $g_s^N g_p^e=0,2\times {10}^{-13},-3\times{10}^{-13}, 4\times {10}^{-13}$ are plotted around $\delta=-\omega_r$ ((a)) and $\delta=\omega_r$  ((b)). Four colors are used, corresponding to the four values of $g_s^N g_P^e$ . (c) The heights of positive (negative) peak in the probe absorption spectrum as a function of $|g_s^N g_P^e|$.   }
\label{fig:3}
\end{figure}

In Fig. 3, we plot the probe absorption spectrums for four values of $g_s^N g_p^e$ around the points of $\delta=-\omega_r$ ((a)) and $\delta=\omega_r$  ((b)). Four different colors are used as shown. It is also seen that when the absolute value of $g_s^N g_p^e$  increases, both the negative and positive peaks become larger. Furthermore, the heights of two peaks are both functions of $|g_s^N g_p^e|$ where ${10}^{-14}\le |g_s^N g_p^e|\le {10}^{-11} $. And the graphs of them which coincide with each other are plotted in Fig. 3(c). Till now, the demonstration of our detection method has been completed. In the following we set a prospective constraint for the coupling constants $g_s^N g_p^e$.

From Fig. 2 we find that in the probe absorption spectrum corresponding to $g=3Hz$ the negative peak at $\delta=-\omega_r$  and the positive peak at $\delta=\omega_r$  are both evident. Based on this, we assume that the minimum value of $g$ could be identified is $0.3Hz$. We also assume that in the relating experiment the exotic monopole-dipole interaction would not be observed. Combining these two assumptions, we can conclude that the value of $g$ corresponding to the experimentally generated probe absorption spectrum would satisfy
\begin{equation}
g<g_m,
\end{equation}
where $g_m \equiv 3Hz$. Then using Eqs. (24)-(25), we could obtain
\begin{equation}
|g_s^N g_p^e|<\frac{2m\gamma g_m}{\lambda |u| g_s \mu_B a_0 \rho},
\end{equation}
where ${10}^{-10}m\le \lambda \le {10}^{-4}m $. Evidently, (26) sets  upper bounds on $g_s^N g_p^e$ as a function of the force range $\lambda$ the domain of which is $[10^{-10}m, 10^{-4}m]$. Consequently, a prospective constraint at ${10}^{-10}m\le \lambda \le {10}^{-4}m $ has been set. And this constraint is presented in Fig. 4.

\begin{figure}
\resizebox{0.5\textwidth}{!}{%
  \includegraphics{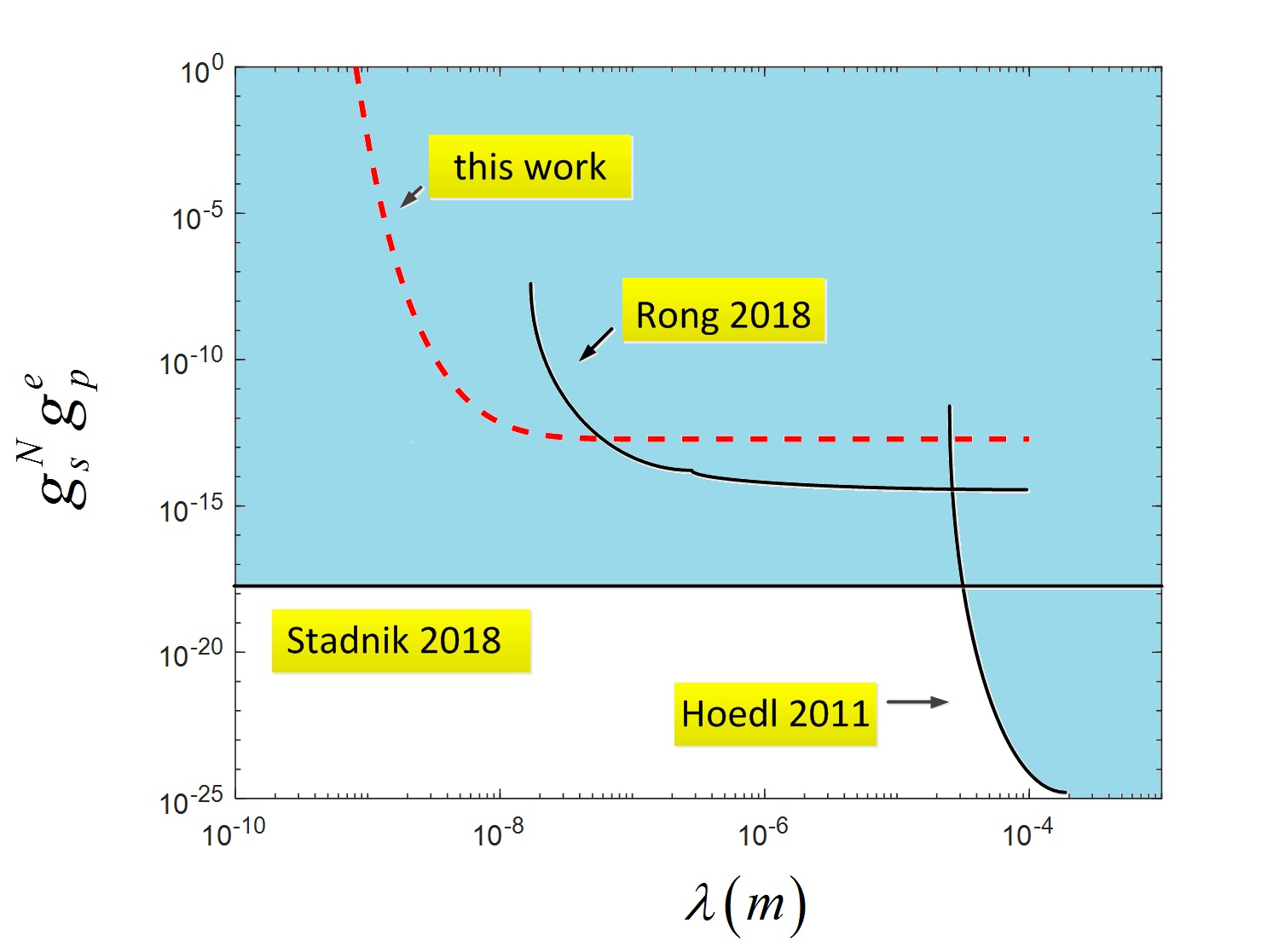}
}
\caption{ Upper limits on $g_s^N g_P^e$  as a function of the force range. Our result is represented as the red dashed curve. Three black solid lines represent the results from Refs. 16, 20 and 31. The pale green region is excluded }
\label{fig:4}
\end{figure}

Now we focus on Fig. 4. Besides our work, three experimental constraints at ultrashort force ranges are also shown in the figure, which are set by Hoedl 2011 \cite{hoedl2011improved}, Rong 2018 \cite{rong2018searching} and Stadnik 2018 \cite{stadnik2018improved} respectively. Differnt methods are used to establish the three constraints. Hoedl et al. utilize a magnetically unshielded torsion pendulum to search for a parity and time-reversal symmetry-violating  force. Rong et al. use a single NV center to detect and constrain the exotic monopole-dipole interaction. Stadnik et al. calculate axion-exchange-induced atomic electric dipole moments (EDMs) including electron core polarization corrections and derive their limit on $g_s^N g_P^e$. Obviously, at the ultrashort force range the constraint of Stadnik is most stringent. And the pale green region is excluded.
\section{\label{sec:level1} Conclusion and outlook}
In summary, we have theoretically proposed a novel method of searching for the electron-nucleon monopole-dipole interaction. Using a hybrid spin-nanocantilever quantum device and applying a static magnetic field and two microwave beams, we could obtain a probe absorption spectrum. For a general specified force range, both the positive peak and the negative one in the absorption spectrum could be considered as the signature of this interaction. Besides, we provide an prospective constraint  for $g_s^N g_P^e$ . Of course, our constraint is only an estimated one and not accurate, and the achievement of the real or right constraint needs relevant experimental search and more theoretical analysis or calculation.

Several points are mentioned here. First, our method deserves consideration in other spin-dependent interactions experimental searches, not limit to the the electron-neucleon monopole-dipole interaction. Second, we can consider  other nanomechanical systems such as nanoparticle and construct relating quantum optical systems to search for hypothetical interactions. Third, it seems that if we perform a hypothetical interaction experimental search in which a quantum optical scheme is employed, the results corresponding to  nanoscale or microscale force range will be most valuable.  Finally, we hope our method would be realized experimentally in the near future.

\begin{acknowledgments}
This work was supported by National Nature Science Foundation of China (11274230.11574206). 
\end{acknowledgments}


\bibliography{apssamp}
\end{document}